\documentclass[aps,pre,twocolumn,superscriptaddress,showpacs,showkeys]{revtex4}

\usepackage{natbib}
\usepackage{dcolumn}
\usepackage{amsmath}
\usepackage{graphicx}
\usepackage{rotating}
\usepackage{multirow}
\begin{document}
\title{Spectral centrality measures in complex networks}

\author{Nicola Perra}
\affiliation{Dipartimento di Fisica, Universit\`a di Cagliari, Italy}
\affiliation{Linkalab, Center for the Study of Complex Networks, Cagliari 09129, Sardegna - Italy}

\author{Santo Fortunato}
\affiliation{Complex Networks Lagrange Laboratory (CNLL),
Institute for Scientific Interchange (ISI), Viale S. Severo 65, 10133, Torino, Italy}

\date{\today} \widetext

\begin{abstract}

Complex networks are characterized by heterogeneous distributions of the degree of nodes, which 
produce a large diversification of the roles of the nodes within the network. Several centrality measures 
have been introduced to rank nodes based on their topological importance within a graph.  
Here we review and compare centrality measures based on spectral properties
of graph matrices. We shall focus on PageRank, eigenvector centrality and the hub/authority scores
of HITS. We derive simple relations between the measures and the (in)degree of the nodes, in some limits. 
We also compare the rankings obtained with different centrality measures. 

\end{abstract}

\maketitle

\section{Introduction}

Complex systems can be represented as networks, 
where the main units of the system become nodes and interacting units are connected by edges.
The last years have witnessed an intense research activity on networks by the scientific community, after the discovery that
many systems in nature, society and technology, turn into graphs with peculiar properties~\cite{Newman:2003, vitorep}.
In particular, many networks are characterized by a heterogeneous distribution of the number of neighbors of a node, or degree,
where nodes with low degree coexist with nodes with large degree (hubs). Such heterogeneity is responsible for a number of 
remarkable features of real networks, such as resilience to random failures/attacks~\cite{albert00}, and the absence of 
a threshold for percolation~\cite{cohen00} and epidemic spreading~\cite{pastor01}. The presence of nodes with different degrees
means that there is a broad diversification of their roles within the graph. Centrality measures are designed to 
rank graph nodes based on their topological importance~\cite{wasserman94,scott00}. 
Among the most popular centrality measures we mention degree itself, but also measures
depending on shortest paths between nodes' pairs, like node betweenness and closeness. 
There are as well centrality measures that depend on spectral properties of graph matrices. These measures   
are important because they are usually associated to simple dynamic processes taking place on graphs, like diffusion.
In particular, the PageRank algorithm, proposed by the Google founders Brin and Page~\cite{brin98}, 
managed to turn Google into the leading interface between users and the World Wide Web.
In recent work spectral properties of graph matrices have also been used to characterize the participation of
nodes in network subgraphs (subgraph centrality)~\cite{estrada1,estrada2} and to estimate the bipartitivity
of graphs~\cite{estrada3}.

However, spectral centrality measures have not been much investigated in the physics literature. 
We shall introduce and review four centrality measures: PageRank, eigenvector centrality~\cite{bonacich01} 
and the hub/authority scores introduced 
by Kleinberg for his HITS algorithm~\cite{kleinberg98}. These measures are usually adopted on directed graphs, 
we shall as well discuss extensions to the undirected case, where applicable.

In Section~\ref{sec1} we present the measures and describe them in some detail.
Analytical and numerical results on particular graphs will be shown in Section~\ref{sec2}, whereas in Section~\ref{sec3} 
we shall compare the rankings of nodes of real graphs for different centrality measures. Conclusions will be  
reported in Section~\ref{sec4}.

\section{Centrality measures}
\label{sec1}

The basic matrix of a graph is the {\it adjacency matrix} $A$, where the element
$A_{ij}$ equals $1$, if nodes $i$ and $j$ are connected by a link, $0$ if they are not. If the network is directed, 
the adjacency matrix is not symmetric. In this case, it is necessary
to distinguish between two types of links adjacent to a node, i. e.
links pointing to the node (incoming) and links pointing outside (outgoing). 
Therefore, there are two types of degree: \textit{indegree}, i.e. the number of incoming links;
\textit{outdegree}, i.e. the number of outgoing links. Likewise, one distinguishes between
the {\it in-neighbors} of a node, i.e. the nodes pointing at the node, and the {\it out-neighbors}, i.e. the nodes pointed at by the node. 
The directedness of the links has a number of important implications, involving 
both some basic structural concepts, like connectivity, and processes taking place on the network. For instance, a random walk is a stationary process 
on any undirected graph, but it is not in general on a directed graph, due to the possible 
presence of {\it dangling ends}, i.e. nodes with zero outdegree,
that act as sinks for the process. 
On the other hand, diffusion leads to a natural definition of centrality, based on the frequency that a walker stops by a node during 
the process. In order to make a diffusive process stationary on a directed graph, one needs to give the walker the opportunity
to leave from a dangling end. PageRank offers a simple solution, which we describe below.

\subsection{PageRank}

PageRank (PR) is the prestige measure used by Google to rank Web pages. It is supposed to simulate the behavior of a user
browsing the Web. Most of the times, the user visits pages just by surfing, i.e. by clicking on hyperlinks of the page
he is on; otherwise, the user will jump to another page by typing its URL on the browser, or going to a bookmark, etc.. On a graph,
this process can be modelled by a simple combination of a random walk with occasional jumps towards randomly selected nodes.
This can be described by the simple set of implicit relations
\begin{equation}
p(i)=\frac{q}{n}+(1-q)\sum_{j: j \rightarrow i}\frac{p(j)}{k_{out}(j)}. \hskip0.7cm i=1,2,
\dots, n
\label{eq1}
\end{equation}
Here, $n$ is the number of nodes of the graph, $p(i)$ is the PR-value of node $i$, $k_{out}(j)$ the outdegree of node
$j$ and the sum runs over the nodes pointing towards $i$. The {\it damping factor} $q$ is a probability, that weighs the mixture
between random walk and random jump. On practical applications it is usually set to small values (typically 0.15).
For any $q>0$ the process reaches stationarity, as a walker has a finite (no matter how small) probability to escape from
a dangling end, whenever it lands there. When $q=0$, the process may not be stationary and PR is ill defined.
When $q=1$, instead, the jumping process dominates and all nodes have the same PR-value $1/n$. PR goes beyond indegree: 
in order to have a large PR-value for a node it is important to have many neighbors pointing at a node, i.e. large indegree, 
but it is also important that the
neighbors have large PR-values. So, if two nodes have equal indegree, the node with more ``important'' neighbors will have 
larger PR.

Solving the set of equations (\ref{eq1}) is equivalent to solving the eigenvalue problem for the 
transition matrix $\cal M$, whose element
${\cal M}_{ij}$ is given by the following expression:
\begin{equation}
{\cal M}_{ij}=\frac{q}{n}+(1-q)\frac{1}{k_{out}(j)}A_{ji}.
\label{eq2}
\end{equation}
PR is just the principal eigenvector of $\cal M$, and is usually determined with the power method, i.e. by
repeatedly multiplying the matrix $\cal M$ by an arbitrary
vector until all the entries of the resulting vector are stable. This is also the procedure we adopted to compute the eigenvectors
corresponding to all centrality measures we studied. 

The literature on PR is very large, because of its huge impact on Web search. In one of the first theoretical studies
~\cite{boldi05}, the dependence of PR on the damping factor was investigated. In general, the attention has been mostly focused on 
the graph of the World Wide Web, where Web pages are nodes and the hyperlinks their connections. 
Comparatively little has been done to study the measure on more general classes of networks. A recent mean field study~\cite{fortunato08} has 
shown that the average PR value of nodes with the same indegree is a linear function of indegree in the 
absence of degree-degree correlations. In another study, some analytical results were found on PR distributions
on special classes of graphs~\cite{fortunato07}. In Section~\ref{subsec3.1} we shall briefly resume the 
results of~\cite{fortunato07} and build up on them. 

\subsection{Eigenvector centrality}

The eigenvector centrality (EV) is also based on the principle that the importance of a node depends on the importance
of its neighbors. In this case the relationship is more straightforward than for PR: the prestige $x_i$ of node $i$ is 
just proportional to the sum of the prestiges of the neighboring nodes pointing to it
\begin{equation}
\lambda x_i=\sum_{j: j \rightarrow i}x_j=\sum_j A_{ji}x_j=({\bf A}^t{\bf x})_i.
\label{eq3}
\end{equation}
From Eq.~(\ref{eq3}) we see that $x_i$ is just the $i$-component of the eigenvector of the transpose of the adjacency matrix with 
eigenvalue $\lambda$. We notice that the trivial eigenvector with all components equal to zero is always a solution of Eq.~(\ref{eq3}).
The true EV is then associated to the existence of non-trivial solutions of the eigenvalue problem of Eq.~(\ref{eq3}).
From Eq.~(\ref{eq3}) we see that
nodes with indegree zero also have zero centrality: in general, nodes pointed at by nodes with zero centrality also have zero centrality
and this effect will propagate to other nodes, so that in many cases EV would not give any information about a big number of nodes. 
To avoid this, it is useful to make the following modification: to each node we assign a prestige $\epsilon$, which is independent of its relationships
with the other nodes. Eq.~(\ref{eq3}) is then modified as follows:
\begin{equation}
x_i=\alpha({\bf A}^t{\bf x})_i+\epsilon.
\label{eq4}
\end{equation}
The role of the parameter $\epsilon$ reminds that of the damping factor $q$ in PR. The parameter $\alpha$ weighs the relative importance
of the contribution of the peers versus that of the node itself. The new measure is called {\it $\alpha$-centrality} ($\alpha$EV)~\cite{bonacich01} 
and is the one we shall investigate in 
this paper. We remark that, in contrast to PR, 
here the solutions do not have a natural interpretation in terms of probability, so the
sum of the $\alpha$-centralities need not be $1$. However we shall normalize the final values by dividing them by their sum, 
so to make them add up to $1$, for practical purposes.

\subsection{HITS scores}

Google's PR was not the first prestige measure for Web pages based on the Web's graph representation. Shortly before the seminal paper
by Brin and Page, Jon Kleinberg~\cite{kleinberg98} had proposed another solution to the problem of ranking 
Web sites based on their importance for the users. This solution was the {\it HITS algorithm}, which distinguishes two types of 
Web pages: {\it hubs} and {\it authorities}. Let us suppose that a user submits a query through a search engine. If a page is very relevant
for this query, one can reasonably expect that it will be pointed at by many other pages. However, 
the simple indegree would not allow to discriminate the
relevant pages from other pages with similar (large) indegree. An important difference is that pages pointing to a relevant page are likely to
point as well to other relevant pages, so to create a sort of bipartite structure where relevant pages (authorities) are cited by special
pages/indices (hubs). Such bipartite structures allow to identify the relevant pages for the user query. Therefore one assigns  
two scores to a page $i$ of the Web: the {\it hub score} $x_i$ and the {\it authority score} $y_i$. Pages with high authority
scores are pointed at by pages with high hub scores. In turn, a good hub points at (very) authoritative pages. 
This mutually reinforcing mechanism is described by the coupled relations
\begin{eqnarray}
\lambda y_i=\sum_{j: j \rightarrow i}x_j=\sum_j A_{ji}x_j=({\bf A}^t{\bf x})_i,\\
\label{eq5}
\mu x_i=\sum_{j: i \rightarrow j}y_j=\sum_j A_{ij}y_j=({\bf A}{\bf y})_i,
\label{eq6}
\end{eqnarray}
which can be rewritten in the form of simple eigenvalue equations for both ${\bf x}$ and ${\bf y}$ by substitution
\begin{equation}
\lambda\mu x_i=({\bf AA^t}{\bf x})_i.
\label{eq7}
\end{equation}
\begin{equation}
\lambda\mu y_i=({\bf A^tA}{\bf y})_i,
\label{eq8}
\end{equation}
From Eqs.~(\ref{eq7}) and ~(\ref{eq8}) we see that the hub and authority scores are just eigenvectors of the matrices
$AA^t$ and $A^tA$. We stress that both $AA^t$ and $A^tA$ are symmetric, whether $A$ is symmetric or not. 
The scores ${\bf x}$ and ${\bf y}$ correspond to the principal eigenvectors of these matrices.

\section{Results}
\label{sec2}

\subsection{PageRank}
\label{subsec3.1}

In~\cite{fortunato07} the two main limits of the PR measure, corresponding to
$q\rightarrow 0$ and $q\rightarrow 1$, were investigated. Analytical results can be derived for special 
graphs, such as graphs grown with popular mechanisms, like preferential attachment~\cite{BA}. For our proofs we shall focus on 
the model by Dorogovtsev, Mendes and Samukhin (DMS)~\cite{DM}, which generates graphs with power-law degree 
distributions with any exponent larger than $2$. In this model, at each time step a new node is added
to the graph and $m$ links are set from the new node to preexisting ones. The probability that a new node $i$ gets attached to a node
$j$ (with indegree $k_j$) is 
\begin{equation}
\Pi(k_j,a)=\frac{a+k_j}{\sum_{l=1}^{i-1}(a+k_l)},
\label{eq9}
\end{equation}
where $a$ is a positive constant. When $a=m$ one recovers the recipe of the original preferential attachment formulation of
Barab\'asi and Albert~\cite{BA}. In general, the exponent of the indegree distribution $\gamma=2+a/m$. For simplicity, we shall study 
the special case in which $m=1$, i.e. each node has outdegree $1$ and the network is a tree. The results are however independent of $m$. 

\subsubsection{The limit $q\rightarrow 0$}
\label{subsubsec3.1}

\begin{figure}[t]
    \includegraphics[width=5cm]{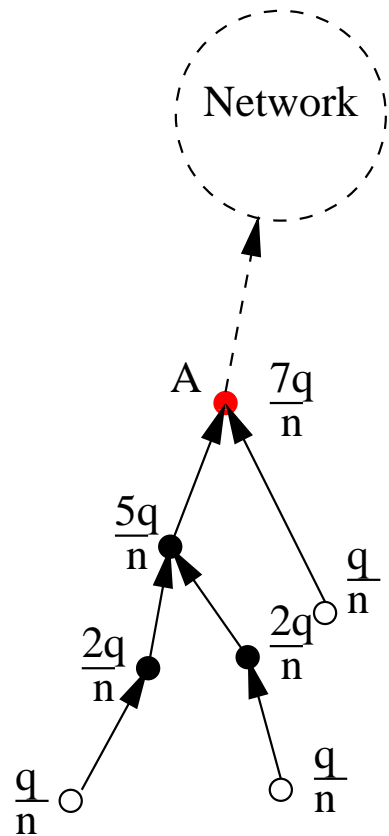}
    \caption{\label{fig1} Subgraph of a tree. The PR-values of all nodes shown can be simply calculated.}
\end{figure}

We assume that $q$ is very small. To the first order in $q$, and remembering that each node has outdegree $1$ by
construction, Eq.~(\ref{eq1}) takes the following form
\begin{equation}
p(i)\sim\frac{q}{n}+\sum_{j: j \rightarrow i}p(j)\hskip0.7cm i=1,2,
\dots, n
\label{eq10}
\end{equation}
which looks particularly simple, though not generally solvable. From Eq.~(\ref{eq10}) we see that the PR of a node equals
a constant plus the PR of its in-neighbors. This recipe enables to calculate PR recursively on simple trees,  
as shown in Fig.~\ref{fig1}, where we focus on a subgraph of a tree.
Node $A$ is the root of the subgraph as every walk starting on any of the nodes will reach $A$ at some stage.
We call any node with this property a {\it predecessor} of $A$. The PR value of any node of the graph is determined only by
its predecessors. In the case illustrated, the calculation is particularly simple: we start from the leaves of the subgraph (empty circles)
whose PR is just $q/n$ because they have no incoming links, and move towards $A$. For each node, we apply the relation (\ref{eq10}).
The final values are reported next to the nodes. From this example we can deduce a number of general properties:
\begin{itemize}
\item{all PR values
are multiples of the elementary unit $q/n$;}
\item{PR increases if one moves from a node to another 
by following a link;}
\item{the PR of each node $i$, in units of $q/n$, 
equals the number of its predecessors.}
\end{itemize}
Since PR takes only discrete values, in the following we shall measure it in units of $q/n$. We thus indicate the distribution
with $P_{PR}(l)$, with $l=1,2,...,n$. 

\begin{figure}[t]
    \includegraphics[width=7.5cm]{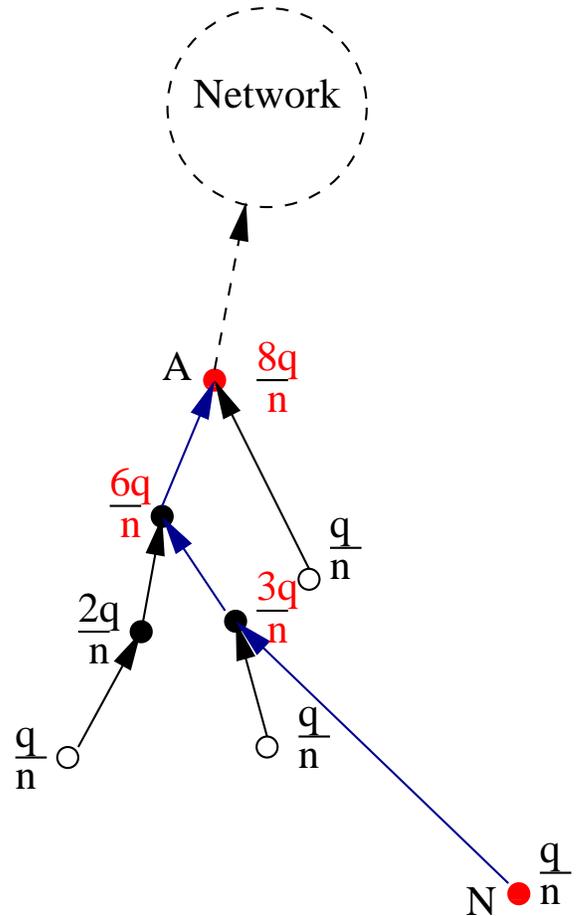}
    \caption{\label{fig2} If a new node N gets attached to any node of the subgraph, it adds an equal contribution
$q/n$ to the PR of all nodes in a path from $N$ to the root.}
\end{figure}

In a dynamic process like network growth, it is crucial to see what happens to the PR values/distribution when a new node
comes into the picture. This is shown in Fig.~\ref{fig2}, where a new node $N$ is added to the network of Fig.~\ref{fig1}. 
We see that only the nodes encountered along the path from $N$ to $A$, including $A$, are affected, while the others retain their
PR values. In particular the presence of the node $N$ determines an increase by $q/n$ in the PR values of the affected nodes.

\begin{figure}[t]
    \includegraphics[width=8cm]{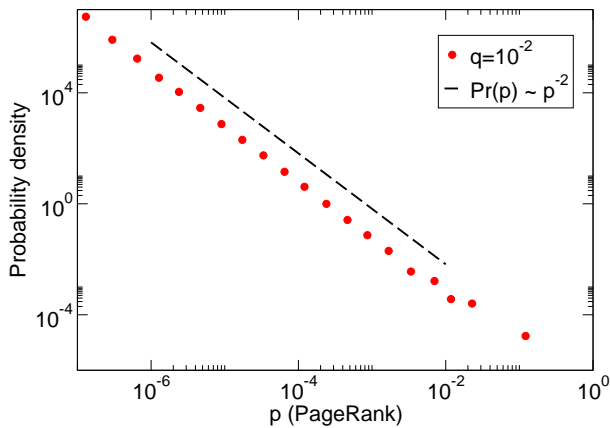}
    \caption{\label{fig3} PR distribution for small $q$ on a DMS graph with $10^6$ nodes, $m=1$ and $a=1$. In this case
the indegree distribution is a power law with exponent $\gamma=3$.}
\end{figure}

Now we are ready to build a master equation for the PR distribution $P_{PR}(l)$ on a DMS graph. At time $n$, the graph has
$n$ nodes and $n-1$ links (the root does not generate links); the PR distribution is ${P_{PR}}^n(l)$. If we add node $n+1$ we get a new distribution 
${P_{PR}}^{n+1}(l)$. As we have seen above, the new node will contribute an additional $q/n$ to the PR of the nodes 
in the path from $n+1$ to the root of the graph. We need to compute the balance between the nodes passing from 
PR $l-1$ to $l$ and those passing from $l$ to $l+1$. 
The probability $\Pi_i^n$ that the PR of node $i$, initially equal to $l$, will be changed by the new node equals the probability that the 
link set by the new node gets attached to one of the predecessors of $i$ (including $i$) and equals 
\begin{equation}
\Pi_i^n=\sum_{j=>i}\frac{a+k_j}{\sum_{t=1}^{n}(a+k_t)}=\sum_{j=>i}\frac{a+k_j}{(a+1)n-1},
\label{eq11}
\end{equation}
where $j=>i$ means that $j$ is a predecessor of $i$. None of the predecessors of $i$, other than $i$ can reach PR $l+1$ because of the
new node, as their initial values are necessarily smaller than $l$. The number of predecessors of $i$ (including $i$) is $l$ and the total number
of adjacent links to the predecessors is $l-1$ (one for each predecessor, except $i$). So,
\begin{equation}
\Pi_i^n=\sum_{j=>i}\frac{a+k_j}{(a+1)n-1}=\frac{(a+1)l-1}{(a+1)n-1}.
\label{eq12}
\end{equation}
The number of nodes with PR $l$ that are affected by the presence of the new node and its link is then
\begin{equation}
\Pi^n(l)=nP_{PR}^{n}(l)\Pi_i^n=\frac{(a+1)l-1}{(a+1)-1/n}P_{PR}^{n}(l).
\label{eq13}
\end{equation}
and the master equation reads
\begin{equation}
(n+1)P_{PR}^{n+1}(l)-nP_{PR}^{n}(l)=\Pi^n(l-1)-\Pi^n(l).
\label{eq14}
\end{equation}
Eq.~(\ref{eq14}) holds for $l>1$. For $l=1$ a modification is necessary, as there cannot be nodes with zero PR, so the term
$\Pi^n(0)$ is not defined. However, since the new node has no incoming links, the number of nodes with PR $1$ 
increases by $1$ because of the new node, so we can write
\begin{equation}
(n+1)P_{PR}^{n+1}(1)-nP_{PR}^{n}(1)=1-\Pi^n(1).
\label{eq15}
\end{equation}
The stationarity condition of Eqs.~(\ref{eq14}) and ~(\ref{eq15}), in the limit of large $n$ leads to the relations
\begin{equation}
P_{PR}(l) = \left\{
  \begin{array}{ll}
\frac{(a+1)l-a-2}{(a+1)l+a}P_{PR}(l-1),
     & \mbox{\hskip1cm   if $l>1$;}\\
    \frac{a+1}{2a+1}, & \mbox{\hskip1cm   if $l=1$.}
  \end{array} \right.
\label{eq16}
\end{equation}
which has the solution
\begin{equation}
P_{PR}(l) = \frac{a(a+1)}{[(a+1)l+a][(a+1)l-1]}\sim \frac{1}{l^2}, \mbox{for $l\gg 1$}.
\label{eq17}
\end{equation}
We see that the PR distribution in the limit $q\rightarrow 0$ on a DMS tree is a power law with exponent $2$, for any value
of the parameter $a$, including the limit case $a\rightarrow\infty$, when the indegree
distribution becomes exponential.
This result is confirmed by numerical simulations (Fig.~\ref{fig3}), which also show that
the hypothesis of the tree is not necessary, as long as each node has the same outdegree $m$. 

In~\cite{fortunato07} the same result was found for other models of network growth, 
like Barab\'asi-Albert preferential attachment~\cite{BA} and the Copying Model~\cite{CM}.
It is possible that this property holds for general graphs where the flows converge towards a central root (sink). Indeed, 
our finding agrees with the more general result on the size distribution of supercritical trees~\cite{delos01}.
Moreover, numerical studies have shown that the same behavior holds for the graph of Internet, when one considers the distribution of 
the size of the basin connected to a given point~\cite{caldarelli00}. Indeed, our calculation follows the same procedure
usually adopted for the calculation of the area of basins in river networks.

\subsubsection{The limit $q\rightarrow 1$}
\label{subsubsec3.2}

\begin{figure}[t]
    \includegraphics[width=8cm]{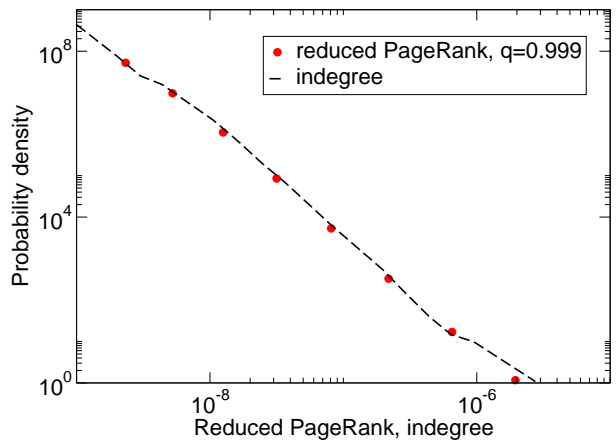}
    \caption{\label{fig4} Reduced PR distribution for $q\sim 1$ on a DMS graph with $10^6$ nodes, $m=1$ and $a=1$. The curve matches the indegree distribution.}
\end{figure}

The case $q=1$ is well defined, but trivial, as all nodes end up having the same PR-value $1/n$.
We ask how this limit is reached. If $q\sim 1$, the contribution to PR given by the in-neighbors of a node 
is very small compared to the constant term, which is close to $1/n$. In order to study the behavior of this term, we define the
{\it reduced PageRank} $p_r(i)$ of a node $i$ as
\begin{equation}
p_r(i)=p(i)-\frac{q}{n}\hskip0.7cm i=1,2,
\dots, n.
\label{eq18}
\end{equation}
We assume that all nodes have the same outdegree $m$. In this case, to leading order in the infinitesimal $1-q$
Eq.~(\ref{eq1}) can be rewritten as
\begin{equation}
p_r(i)=\frac{q(1-q)}{mn}k_{in}(i),\hskip0.7cm i=1,2,
\dots, n. 
\label{eq19}
\end{equation}
where $k_{in}(i)$ is the indegree of $i$. We conclude that on any graph, the reduced PR of a node in the limit
$q\rightarrow 1$ is proportional to the indegree of the node, if all nodes have the same outdegree. 
This result has been derived independently in~\cite{chen07}. As a consequence of Eq.~(\ref{eq19}), the distribution of the 
reduced PR for $q\rightarrow 1$ has the same trend as that of indegree, which can be easily verified numerically (Fig.\ref{fig4}). 

\medskip

\subsubsection{Extension to undirected graphs}
\label{subsubsec3.3}

\begin{figure}[t]
    \includegraphics[width=7.8cm]{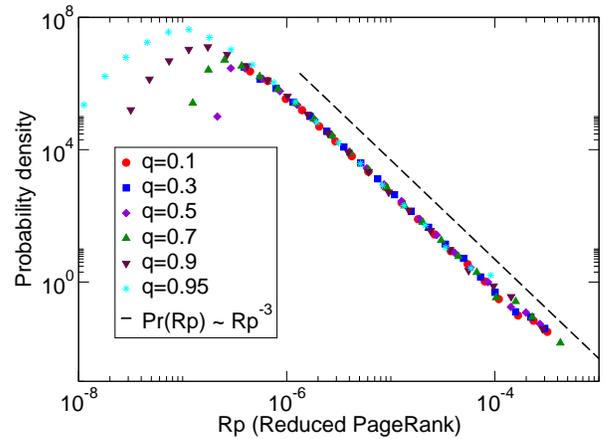}
    \caption{\label{fig5} Reduced PR on undirected graphs. Variability of reduced PR 
distribution with $q$ on a DMS graph with $10^6$ nodes, $m=3$ and $a=3$. The
degree distribution has a power law tail with exponent $\gamma=3$.}
\end{figure}

\begin{figure}[t]
    \includegraphics[width=7.8cm]{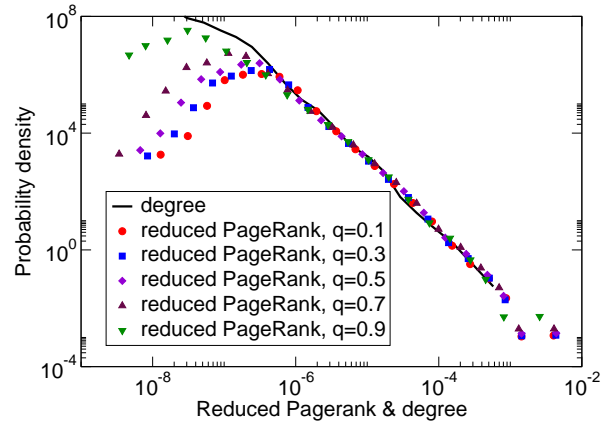}
    \caption{\label{fig5a} Reduced PR on undirected graphs. Variability of reduced PR 
distribution with $q$ on the domain {\tt .gov} of the World Wide Web. The degree distribution has a tail which 
follows fairly well a power law with exponent $2.1$. To better show the agreement  
we have shifted the curves such that the tails overlap.}
\end{figure}

PR can be easily extended to undirected graphs as well. The corresponding equation reads
\begin{equation}
p(i)=\frac{q}{n}+(1-q)\sum_{j: j \leftrightarrow i}\frac{p(j)}{k_j}. \hskip0.7cm i=1,2,
\dots, n
\label{eq20}
\end{equation}

where now $k_j$ is the degree of node $j$. For the purposes of a random walk, undirected links can be 
crossed in both directions, so a pure random walk now always reaches stationarity due to the absence of dangling ends.
In fact, the stationary probability of a random walk on a node of any undirected graph is simply 
proportional to the degree of the node~\cite{bollobas}.
However, in Eq.~(\ref{eq20}) we have still the contribution of random jumping, and it turns out that the mixed process
is still hard to solve. We are not aware of a general solution in this case. In the limit $q\rightarrow 0$ PR is now
well behaved, and its distribution coincides with the degree distribution of the graph. In Fig.~\ref{fig5} we show the 
distributions of reduced PR for different values of $q$ on a DMS graph with a power law degree distribution and exponent $\gamma=3$.
The reduced PR expresses the contribution to PR given by the random walk.
We see that the curves follow the decay of the degree distribution
for any value of $q$. We have computed the reduced PR distribution on many other graphs and in all cases we found that 
they follow the same trend as the degree distribution. For example, in Fig.~\ref{fig5a} we show the comparison between 
reduced PR and degree for a sample of the Web link graph. Here the nodes are Web pages
of the domain {\tt .gov} and two pages are connected if there is a hyperlink from one to the other. There are $794,184$ nodes and 
$6,460,903$ links. The graph is directed
but PR was calculated by neglecting the directedness of the links. As we can see, the decay of the distributions of reduced PR 
resembles that of the degree distribution. The graph at hand is not simple like the DMS networks, as it presents a large number of loops
and community structure. Therefore the result is likely to be general. We can show this with a simple argument. 
The general equation for reduced PR on undirected graphs is:

\begin{equation}
p_r(i)=\frac{(1-q)q}{n}\sum_{j: j \leftrightarrow i}\frac{1}{k_j}+(1-q)\sum_{j: j \leftrightarrow i}\frac{p_r(j)}{k_j},
\label{eq21}
\end{equation}
that we can solve formally by successive iteration, obtaining the general form
\begin{equation}
\begin{split}
p_r(i)=\frac{q}{n}\sum_{s}(1-q)^s\sum_{i_1}\frac{1}{k_{i_1}}\sum_{i_2}\frac{1}{k_{i_2}}...\sum_{i_s}\frac{1}{k_{i_s}}\\
=\frac{q}{n}\sum_{s}(1-q)^s \prod_{i_1\leftrightarrow i_2...\leftrightarrow i_s}\frac{1}{k_{i_s}},
\end{split}
\label{eq22}
\end{equation}
where $i_s$ indicates the neighbors of the $s$-shell of the node $i$; so, $i_1$ indicates the nearest
neighbors of $i$, $i_2$ the next-to-nearest neighbors, and so on. The last sum in the first line of Eq.~(\ref{eq22})
is, for a given node $i_{s-1}$, a sum over its neighbors $i_s$. This sum, that we call $T_{i_s}$, contains $k_{i_s}$ terms, $k_{i_s}$ being
the degree of node $i_s$. 
The sum $T_{i_s}$ can be approximated as the product $k_{i_s}\langle 1/k\rangle_{NN}$, where $\langle 1/k\rangle_{NN}$ is the expected
value of the average of $1/k$ over the neighbors of a node of the network.  
In general, $T_{i_s}=k_{i_s}\langle 1/k\rangle_{NN}+\eta_{i_s}$, where $\eta_{i_s}$ is a random variable with mean zero. 
In this way, it is easy to see from Eq.~(\ref{eq22}) that, for any value of $s$, the product of sums reduces to $k_{i}\langle 1/k\rangle_{NN}$ plus 
the sum of many random variables like $\eta_{i_s}$. Due to the Central Limit Theorem, the latter sum, if it includes a large number of terms,
yields a very small value with large probability. We can then conclude that, for $k_i$ sufficiently large, each term of the
series in Eq.~(\ref{eq22}) is proportional to $k_i$ with good approximation, therefore $p_r(i)$ is also proportional to $k_i$, for any 
value of the damping factor $q$.  
We have verified 
numerically that this assertion is true for many graphs and degree distributions, without finding exceptions. 

\medskip
\medskip

\subsection{Eigenvector centrality}
\label{subsec3.2}

\subsubsection{Directed graphs}
\label{subsubsec3.4}

\begin{figure}[t]
    \includegraphics[width=8cm]{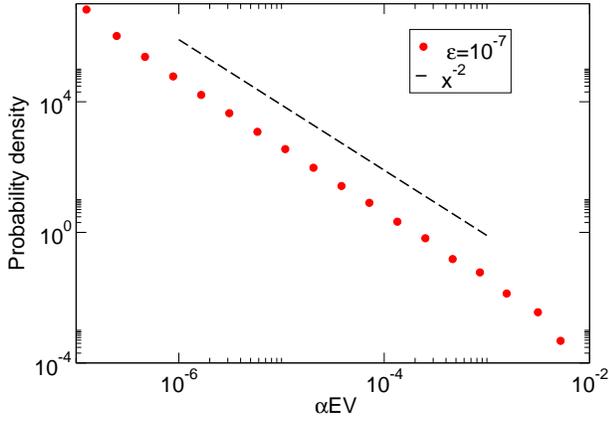}
    \caption{\label{fig6} Distribution of $\alpha$EV on a directed DMS graph with $10^6$ nodes, $m=1$ and $a=1$. The dashed line 
indicates the predicted slope.}
\end{figure}

The defining Eq.~(\ref{eq4}) is formally analogous to Eq.~(\ref{eq10}). The only difference is that the
eigenvalue $\alpha$ is not $1$ as for PR. However, the results of Section ~\ref{subsubsec3.1} hold as  
well when the outdegree $m$ is greater than $1$ (as long as it is the same for all nodes), 
and in this case the sum of Eq.~(\ref{eq10}) would include a multiplicative
factor $1/m$, which makes it identical to Eq.~(\ref{eq4}).
We then deduce that all results found for 
PR in the limit $q\rightarrow 0$ hold for $\alpha$EV. Here the results are more general, because we did not need to
make any approximation to get to Eq.~(\ref{eq4}) as we instead needed to derive Eq.~(\ref{eq10}). In particular, it is not necessary that 
$\epsilon$ be very small and
the nodes need not have the same outdegree, although this is the case for the graphs we considered.
We conclude that the distribution of $\alpha$EV on DMS graphs has a power law tail 
with exponent $2$ (Fig.~\ref{fig6}). The same holds for graphs built 
using preferential attachment and the Copying Model, just as it happens 
for PR in the limit $q\rightarrow 0$. 

\medskip

\subsubsection{Extension to undirected graphs}
\label{subsubsec3.5}

\begin{figure}[t]
    \includegraphics[width=7.8cm]{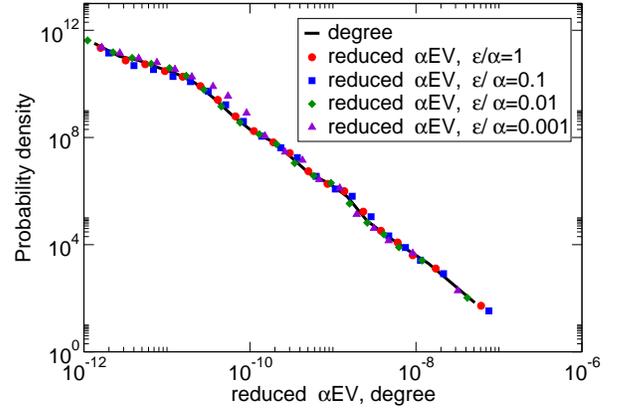}
    \caption{\label{fig6_0} Reduced $\alpha$EV on undirected graphs. Variability of reduced $\alpha$EV 
distribution with $\epsilon/\alpha$ on the domain {\tt .gov} of the World Wide Web. The degree distribution has a tail which 
follows fairly well a power law with exponent $2.1$. To better show the agreement  
we have shifted the curves such that the tails overlap.}
\end{figure}

On undirected graphs, Eq.~(\ref{eq4}) becomes
\begin{equation}
x_i=\alpha({\bf A}{\bf x})_i+\epsilon,
\label{eq23}
\end{equation}
since $A^t=A$. So, the $\alpha$EV of a node is proportional to the sum of the $\alpha$EV
of its neighbors, modulo an additive constant $\epsilon$. As we have done for PR, 
we define the {\it reduced $\alpha$-centrality} as
\begin{equation}
x^r_i=x_i-\epsilon.
\label{eq24}
\end{equation}
So, we can rewrite Eq.~(\ref{eq24}) as
\begin{equation}
x^r_i=\alpha({\bf A}{\bf x^r})_i+k_i\alpha\epsilon,
\label{eq25}
\end{equation}
where $k_i$ is again the degree of node $i$.
We can apply a similar argument as in Section~\ref{subsubsec3.3}. The sum over the $k_i$ neighbors of $i$ can be
approximated as $k_i\langle x^r\rangle$, where $\langle x^r\rangle$ is the average of the reduced $\alpha$EV
over the whole graph. The approximation is the more valid, the larger the number $k_i$ of summands. In this way, from Eq.~(\ref{eq25})
we see that the reduced $\alpha$EV of a node is proportional to its degree, if the latter is large enough. This result is independent
of the specific graph we consider, and we have verified it numerically for many types of networks. In Fig.~\ref{fig6_0} we show the distribution of 
reduced $\alpha$EV for different choices of the parameter $\epsilon/\alpha$ 
for the sample of the Web graph we analyzed in Fig.~\ref{fig5a}. The curves closely follow the decay
of the degree distribution. 

\medskip

\subsection{HITS scores}
\label{subsec3.3}

\begin{figure}[t]
    \includegraphics[width=7cm]{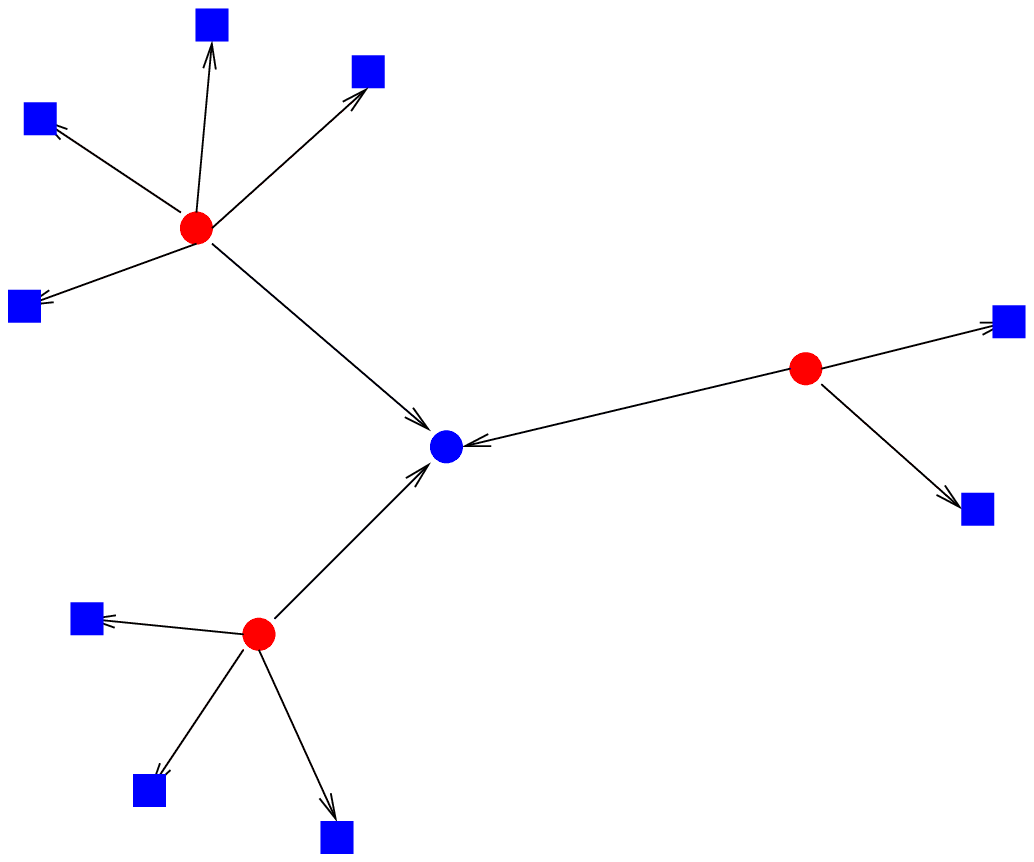}
    \caption{\label{fig6a} The authority score of the node in the center is proportional 
to the sum of the authority scores of the out-neighbors (blue squares)
of the in-neighbors (red circles) of the node.}
\end{figure}

\begin{figure}[t]
    \includegraphics[width=8cm]{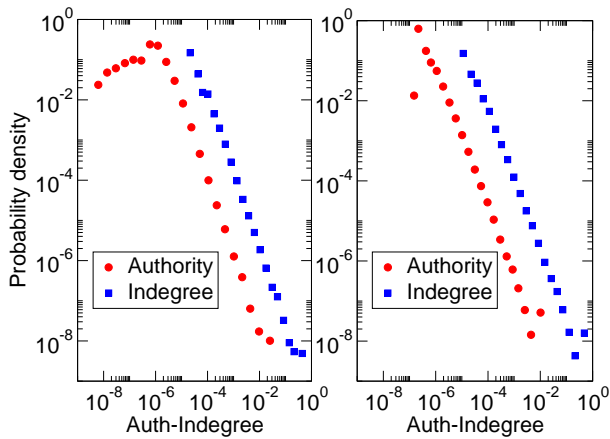}
    \caption{\label{fig7} Distribution of the authority scores versus indegree distribution. (Left) DMS graph with $10^5$ nodes, 
$m=10$ and $a=1$. (Right) DMS graph with $10^5$ nodes, $m=50$ and $a=1$. }
\end{figure}

The meaning of the eigenvalue equations (\ref{eq7}) and (\ref{eq8}) is quite simple. The hub score
of a node is the sum of the hub scores of the in-neighbors of the out-neighbors of the node. The authority score 
of a node is the sum of the authority scores of the out-neighbors of the in-neighbors of the node
(Fig.~\ref{fig6a}).
Let us suppose that the nodes have the same
outdegree $m$. The authority score of a node $i$ is given by the sum of $mk_{in}(i)$ terms, where $k_{in}(i)$ is the indegree of $i$.
In fact, node $i$ has $k_{in}(i)$ in-neighbors, each of them having $m$ out-neighbors.
If $k_{in}(i)$ is large, the number of summands is very large, and can be approximated 
by the average value of the authority score over the whole graph,
times $mk_{in}(i)$. This approximation is the more valid, the larger $m$ and $k_{in}(i)$. We conclude that on a directed graph with constant outdegree
the distribution of the authority scores will have the same tail as the indegree distribution. This is clearly
illustrated in Fig.~\ref{fig7}. For the hub scores it is not possible to make predictions; the sum that delivers
the hub score of a node cannot be approximated through other graph variables in most cases.

The extension of the HITS scores to the case of undirected graphs is not interesting. In this case $A^t=A$, so $A^tA=AA^t=A^2$ and the hub 
and authority scores are identical. Moreover, they coincide with EV, as the matrices
$A$ and $A^2$ have the same eigenvectors.

\medskip

\section{Rankings}
\label{sec3}

In the previous sections we have investigated the distributions of spectral centrality measures and their similarities.
As we have mentioned in the Introduction, centrality measures are used to rank nodes. In this section we shall compare the rankings
obtained with different centrality measures. In order to compare two rankings we adopt Kendall's $\tau$ ~\cite{kendall38}, a widely used 
index in this type of analysis. Kendall's $\tau$ ranges from $1$ (perfect correlation) to $-1$ (perfect anticorrelation).  
In Table ~\ref{tab1} we show the cross-comparisons between all centrality measures we discuss in this 
work, for a DMS directed graph. For completeness we have included the outdegree as well. As we can see, PR, $\alpha$EV and the authority scores
are well correlated with indegree and with each other, whereas the other coefficients are small or negative; $\alpha$EV has a strong correlation
with outdegree as well.
\begin{table}[htbp]
\begin{center}

\begin{tabular}{|c|r|}
\hline
Measures &  $\tau$ \\ 
\hline
\hline
PR-$\alpha$EV & 0.8192\\ 
\hline
PR-AUTH &  0.5774\\ 
\hline
PR-HUBS &  0.1213\\
\hline 
PR-IN &  0.6444\\
\hline 
PR-OUT &  -0.3012 \\
\hline 
$\alpha$EV-AUTH &  0.5788 \\
\hline 
$\alpha$EV-IN &   0.6487 \\
\hline 
$\alpha$EV-HUBS & 0.1220\\
\hline
$\alpha$EV-OUT &  0.5788\\
\hline
 AUTH-IN &  0.5458\\
\hline
 AUTH-HUBS & 0.1076 \\
\hline 
AUTH-OUT & -0.2611\\
\hline 
HUBS-IN & 0.1142 \\
\hline
HUBS-OUT & -0.2126\\
\hline
IN-OUT & -0.2507\\
\hline
\end{tabular}
\end{center}
\caption{Kendall's $\tau$ for each pair of centrality measures computed for a DMS directed graph, with $n=10^{6}$, $m=3$ and $a=3$.}
\label{tab1}
\end{table} 

DMS graphs have a fairly regular structure; we have seen that in this case the behavior of centrality measures is quite regular, 
and that there are simple relations between their distributions, which may be determined by simple relations between 
a measure and indegree at the level of the single node. Therefore,
we cannot deduce general conclusions from Table~\ref{tab1} and we repeated the analysis
for two real world networks: a network of political blogs and the subset of the Web link graph corresponding to the URLs of
the domain {\tt .gov}, that we have studied in the previous sections. 

The first network is a citation network consisting of $1490$ blogs; $758$ are democratic and $732$ republican.
It was first studied by Adamic and Glance~\cite{adamic05}, who focused on the community structure of the graph, which 
matches that determined by the two political areas.
\begin{table}[htbp]
\begin{center}
\begin{tabular}{|c|r|}
\hline
Measures &  $\tau$ \\ 
\hline
\hline
PR-$\alpha$EV &  0.09 \\ 
\hline
PR-AUTH & 0.14 \\ 
\hline
PR-HUBS &  0.04\\
\hline 
PR-IN &  0.14 \\
\hline 
PR-OUT & 0.02 \\
\hline 
$\alpha$EV-AUTH & 0.12 \\
\hline 
$\alpha$EV-IN & 0.07 \\
\hline 
$\alpha$EV-HUBS & 0.08\\
\hline
$\alpha$EV-OUT &  0.01 \\
\hline
 AUTH-IN & 0.12 \\
\hline
 AUTH-HUBS & 0.07 \\
\hline 
AUTH-OUT & 0.01 \\
\hline 
HUBS-IN & 0.02 \\
\hline
HUBS-OUT & 0.07  \\
\hline
IN-OUT & 0.07 \\
\hline
\end{tabular}
\end{center}
\caption{Kendall's $\tau$ for each pairs of centrality
measures for the network of political blogs studied by Adamic and Glance.}
\label{tab2}
\end{table}
The correlations now are rather weak. The small coefficients
indicate that the rankings differ considerably with the measure chosen. To have an idea, in Table~\ref{tab3} we show
the Top Ten blogs in the rankings obtained with all centrality measures. We see that there are clear differences between the listings.
\begin{table*}%[htbp]
\centering
\footnotesize
\begin{tabular}{|r|r|r|r|}
\hline
Rank & PR & $\alpha$EV & Auth  \\ 
\hline
\hline
$1^{\circ}$ & dailykos.com, D  & atrios.blogspot.com, D & dailykos.com, D\\ 
\hline
$2^{\circ}$ &  atrios.blogspot.com, D & dailykos.com, D & talkingpointsmemo.com, D \\ 
\hline
$3^{\circ}$ &  instapundit.com, R & talkingpointsmemo.com, D & atrios.blogspot.com, D \\
\hline 
$4^{\circ}$ &  blogsforbush.com, R &  washingtonmonthly.com, D  & washingtonmonthly.com, D\\
\hline 
$5^{\circ}$ &   talkingpointsmemo.com, D &   talkleft.com, D      & talkleft.com, D \\
\hline 
$6^{\circ}$ &  michellemalkin.com, R & prospect.org/weblog, D & instapundit.com, R \\
\hline 
$7^{\circ}$ &  drudgereport.com, R & juancole.com, D  &  juancole.com, D \\
\hline 
$8^{\circ}$ &  washingtonmonthly.com, D & digbysblog.blogspot.com, D &  yglesias.typepad.com/matthew, D \\
\hline 
$9^{\circ}$ & powerlineblog.com, R & pandagon.net, D & pandagon.net, D  \\
\hline
$10^{\circ}$ &  andrewsullivan.com, R & yglesias.typepad.com/matthew, D & digbysblog.blogspot.com, D \\
\hline
\hline
Rank & Hubs & In & Out \\
\hline
\hline
$1^{\circ}$& politicalstrategy.org, D & dailykos.com, D& blogsforbush.com, R\\ 
\hline
$2^{\circ}$ &   madkane.com/notable.html, D & instapundit.com, R& newleftblogs.blogspot.com, D\\ 
\hline
$3^{\circ}$ &  liberaloasis.com, D & talkingpointsmemo.com, D & politicalstrategy.org, D\\
\hline 
$4^{\circ}$ &   stagefour.typepad.com/commonprejudice, D  & atrios.blogspot.com, D& madkane.com/notable.html, D\\
\hline 
$5^{\circ}$ &   bodyandsoul.typepad.com , D & drudgereport.com, R & cayankee.blogs.com, R\\
\hline 
$6^{\circ}$ &    corrente.blogspot.com, D & powerlineblog.com, R& liberaloasis.com, D\\
\hline 
$7^{\circ}$ &   aurelientt.blogspot.com, D  & blogsforbush.com, R & lashawnbarber.com, D\\
\hline 
$8^{\circ}$ &  tbogg.blogspot.com, D     & washingtonmonthly.com, D & gevkaffeegal.typepad.com/thealliance, R\\
\hline 
$9^{\circ}$ &  newleftblogs.blogspot.com, D&   michellemalkin.com, R& presidentboxer.blogspot.com, R \\
\hline
$10^{\circ}$ &  atrios.blogspot.com, D &truthlaidbear.com, R& corrente.blogspot.com, D \\
\hline
\end{tabular}
\caption{Top Ten of the network of political blogs according to PR, $\alpha$EV, authorities, hubs, indegree and outdegree. 
\textbf{D} democratic, \textbf{R}, republican.}
\label{tab3}
\end{table*} 

The results are basically the same for the Web graph. Table~\ref{tab4} reports the Kendall's $\tau$ between
the rankings. The values are of the same magnitude as for the network of the blogs.
The Top Ten listings for the Web are shown in Table~\ref{tab5} and appear again considerably different from each other.
\begin{table}[htbp]
\begin{center}
\begin{tabular}{|c|r|}
\hline
Measures &  $\tau$ \\ 
\hline
\hline
PR-$\alpha$EV &  0.189 \\ 
\hline
PR-AUTH & 0.079 \\ 
\hline
PR-HUBS &  0.060\\
\hline 
PR-IN &  0.155 \\
\hline 
PR-OUT & 0.090 \\
\hline 
$\alpha$EV-AUTH & 0.081 \\
\hline 
$\alpha$EV-IN & 0.147 \\
\hline 
$\alpha$EV-HUBS & 0.074\\
\hline
$\alpha$EV-OUT &  0.086 \\
\hline
 AUTH-IN & 0.046 \\
\hline
 AUTH-HUBS & 0.109 \\
\hline 
AUTH-OUT & 0.072 \\
\hline 
HUBS-IN & 0.003 \\
\hline
HUBS-OUT & 0.056  \\
\hline
IN-OUT & 0.081 \\
\hline
\end{tabular}
\end{center}
\caption{Kendall's $\tau$ for each pairs of centrality
measures for the domain {\tt .gov} of the Web.}
\label{tab4}
\end{table}

\begin{table*}
\centering
\footnotesize
\begin{tabular}{|r|r|r|}
\hline
Rank & PR & $\alpha$EV   \\ 
\hline
\hline 
$1^{\circ}$ &  www.usgs.gov  & polar.wwb.noaa.gov$/$waves$/$main\_int.js \\
\hline
$2^{\circ}$ &   www.nws.noaa.gov & polar.wwb.noaa.gov$/$waves$/$welcome.html \\
\hline
$3^{\circ}$ & www.naca.larc.nasa.gov$/$readme.html  & polar.wwb.noaa.gov$/$waves$/$main\_table.html\\ 
\hline 
$4^{\circ}$ &  www.usda.gov &  polar.wwb.noaa.gov$/$waves$/$products.html \\
\hline 
$5^{\circ}$ &  www.nws.noaa.gov$/$disclaimer.html  &  polar.wwb.noaa.gov$/$waves$/$main\_int.html   \\
\hline 
$6^{\circ}$ & www.ar.inel.gov$/$home.htm &  www.nws.noaa.gov$/$disclaimer1.html\\
\hline 
$7^{\circ}$ & www.4woman.gov$/$search$/$search.cfm   &  www.nws.noaa.gov \\
\hline 
$8^{\circ}$ & www.nws.noaa.gov$/$feedback.shtml & polar.wwb.noaa.gov$/$waves$/$references.htm \\
\hline 
$9^{\circ}$ & www.access.wa.gov & polar.wwb.noaa.gov$/$waves$/$validation.htm \\
\hline
$10^{\circ}$ &  www.usinfo.state.gov$/$products$/$pdq$/$pdq.htm & polar.wwb.noaa.gov$/$waves$/$valid\_wna.html  \\
\hline
\hline
Rank & Auth & In  \\
\hline
\hline
$1^{\circ}$&  www.srh.noaa.gov$/$oun$/$cgi-bin$/$wxclick.pl?county=oklahoma & www.usgs.gov\\
\hline
$2^{\circ}$ & www.srh.noaa.gov$/$oun$/$cgi-bin$/$wxclick.pl?county=cleveland   & www.cdc.gov\\ 
\hline
$3^{\circ}$ & www.srh.noaa.gov$/$oun$/$cgi-bin$/$wxclick.pl?county=kiowa & www.usda.gov \\
\hline 
$4^{\circ}$ &   www.nws.noaa.gov  & www.doi.gov \\
\hline 
$5^{\circ}$ &   www.srh.noaa.gov$/$oun$/$cgi-bin$/$wxclick.pl?county=logan & www.nws.noaa.gov \\
\hline 
$6^{\circ}$ &  www.srh.noaa.gov$/$oun$/$cgi-bin$/$wxclick.pl?county=payne  & www.usgs.gov$/$disclaimer.html\\
\hline 
$7^{\circ}$ &  www.srh.noaa.gov$/$oun$/$cgi-bin$/$wxclick.pl?county=knox  & www.usda.gov$/$news$/$privacy.htm\\
\hline 
$8^{\circ}$ &   weather.noaa.gov   & www.abag.ca.gov \\ 
\hline 
$9^{\circ}$ &  weather.noaa.gov$/$weather$/$ok\_cc\_us.html& www.ars.usda.gov$/$nodisc.html \\
\hline
$10^{\circ}$ &   www.crh.noaa.gov$/$ddc &  www.ars.usda.gov$/$comm.htm\\
\hline
\end{tabular}
\caption{Top ten of the web domain {\tt .gov} according to PR, $\alpha$EV, authorities and indegree.}
\label{tab5}
\end{table*}

\section{Conclusions}
\label{sec4}

Centrality measures are very important to understand the properties of the nodes of complex networks and their topological roles.
We have studied the most important centrality measures based on properties of graph matrices: PageRank, Eigenvector centrality,
and the hub and authority scores of HITS. All these measures deduce the importance of a node in a self-consistent way 
from the importance of its nearest neighbors and, in the case of the HITS scores, of its next-to-nearest neighbors.
We have resumed some recent results on PageRank distributions on particular types of tree-like graphs. On those graphs, the
distribution of PageRank in the limit $q\rightarrow 0$ decays as a power law with exponent $2$. The same is true for $\alpha$-centrality,
because its defining equation is formally equivalent to the equation for PageRank in the limit $q\rightarrow 0$. These results on centrality 
distributions are likely to be true for an extended class of graphs, where there is a flow from the outermost nodes (leaves) to a sink.
We have also seen that, on any graph, in the limit
$q\rightarrow 1$, the reduced PageRank of a node, i.e. the contribution of the random walk process to the measure, is simply proportional to the indegree
of the node, if the nodes have (about) the same outdegree. We have studied for the first time the extension of PageRank to 
the case of undirected networks, finding that the reduced PageRank of a node is proportional to its degree, for large degrees, for any graph and value of $q$. 
We proposed a simple explanation of this effect based on the Central Limit Theorem, and verified numerically in several cases that the argument holds. 
Similarly, the reduced $\alpha$-centrality of a node is also proportional to its degree, for large degrees, on any graph. With the same type of argument
it is possible to show that the authority score of a node is proportional to its indegree, for large indegrees, when the outdegrees of all nodes
are (approximately) the same. 

We conclude that there are often strong relations between our centrality measured and (in)degree: some relations hold on particular graphs and/or limits,
others are more general. These findings imply that the measures are often strongly correlated with each other. We have indeed seen that the rankings
of nodes according to the centrality measures we have considered are quite close to each other for indegree, PageRank, Eigenvector centrality 
and authority score on graphs built with the prescription of Dorogovtsev, Mendes and Samukhin. We have shown in the paper that these graphs
have special properties, and that some measures may be correlated to each other. Instead, on real graphs, like the networks
of political blogs and the sample of the Web graph we have considered, the structure is less regular and the measures are far less correlated
to each other, as confirmed by the small values of the Kendall's $\tau$ for each pair of centrality measures. This means that, for practical purposes, 
and in spite of their similarities, spectral centrality measures look at nodes from different perspectives, and allow to diversify
their roles within the network, obtaining in this way more information about the importance of nodes. The scores computed from
spectral centrality measures can 
complement the information about node's centrality derived from more traditional measures like node betweenness~\cite{freeman}. This is especially important
for directed graphs, where node betweenness, as well as other measures based on geodesic paths, like closeness~\cite{sabidussi}, are not well defined.

\begin{acknowledgments}
We thank A. Lancichinetti, F. Radicchi and A. Vespignani for interesting discussions.
NP thanks the ISI Foundation for support and hospitality
during the project.

\end{acknowledgments}


\begin{thebibliography}{20}

\bibitem{Newman:2003} M.~E.~J. Newman,
SIAM Rev. {\bf 45}, 167 (2003). 

\bibitem{vitorep} 
S. Boccaletti, V. Latora, Y. Moreno, M. Chavez and D.-U. Hwang,
Phys. Rep. {\bf 424}, 175 (2006).

\bibitem{albert00}
R. Albert, H. Jeong and A.-L. Barab\'asi,
Nature {\bf 406}, 378 (2000).

\bibitem{cohen00}
R. Cohen, K. Erez, D. ben Avraham and S. Havlin,
Phys. Rev. Lett. {\bf 85}, 4626 (2000).

\bibitem{pastor01}
R. Pastor-Satorras and A. Vespignani,
Phys. Rev. Lett. {\bf 86}, 3200 (2001).

\bibitem{wasserman94}
S. Wasserman and K. Faust,
{\it Social Networks Analysis}, Cambridge University Press, Cambridge, UK (1994).

\bibitem{scott00}
J. Scott, {\it Social Networks Analysis: a Handbook}, Sage Publications, London (2000).

\bibitem{brin98} S. Brin and L. Page, 
Comput. Netw. {\bf 30}, 107 (1998).

\bibitem{estrada1} E.~Estrada and J.~A.~Rodr\'iguez-Vel\'azquez, Phys. Rev. E {\bf 71}, 056103 (2005).

\bibitem{estrada2} E.~Estrada and N.~Hatano, Chem. Phys. Lett. {\bf 439}, 247 (2007).

\bibitem{estrada3} E.~Estrada and J.~A.~Rodr\'iguez-Vel\'azquez, Phys. Rev. E {\bf 72}, 046105 (2005).

\bibitem{bonacich01} P. Bonacich and P. Lloyd, Soc. Netw. {\bf 23}, 191 (2001).

\bibitem{kleinberg98} J. Kleinberg,
Proc. Ninth Ann. ACM-SIAM Symp. Discrete Algorithms, 668-677, ACM Press, New York (1998).

\bibitem{boldi05} P. ~Boldi, M.~Santini and S.~Vigna, 
\textit{Proc. of the Fourteenth International World Wide Web Conference}, 
Chiba, Japan, ACM Press pp. 557-566 (2005).

\bibitem{fortunato08} S. Fortunato, M. Bogu\~n\'a, A. Flammini and F. Menczer, Internet Math., to appear (2008),
e-print arxiv:cs/0511016.
 
\bibitem{fortunato07} S. Fortunato and A. Flammini, Int. J. Bif. Ch. {\bf 17}, 2343 (2007).

\bibitem{BA} A.-L. Barab{\'a}si and R.~Albert,  
Science \textbf{286}, 509 (1999).

\bibitem{DM} S.~N. Dorogovtsev, J.~F.~F. Mendes and A.~N.~Samukhin, 
Phys. Rev. Lett. {\bf 85}, 4633 (2000).

\bibitem{CM} J. Kleinberg, S.~R. Kumar, P. Raghavan, S. Rajagopalan and A. Tomkins,  
Lect. Notes. Comp. Sci. {\bf 1627}, 1 (1999).

\bibitem{delos01} P.~De Los Rios, Europhys. Lett. {\bf 56}, 898 (2001).

\bibitem{caldarelli00} G.~Caldarelli, R.~Marchetti and L.~Pietronero, Europhys. Lett. {\bf 52}, 386 (2000).

\bibitem{chen07} P. Chen, H. Xie, S. Maslov and S. Redner,
J. Informet. {\bf 1}, 8 (2007).

\bibitem{bollobas} B. Bollob{\'a}s, {\it Modern Graph Theory}. Springer Verlag, New York, USA (1998).

\bibitem{kendall38} M. Kendall, Biometrika {\bf 30}, 81 (1938).

\bibitem{adamic05} L. Adamic and N. Glance, 
{\it Proc. $3^{rd}$ Int. Workshop
on Link Discovery} (Information Sciences Inst., Univ of Southern California, Los Angeles), p. 36 (2005).

\bibitem{freeman} 
L.~C.~Freeman, Sociometry {\bf 40}, 35 (1977).

\bibitem{sabidussi}
G. Sabidussi, Psychometrika {\bf 31}, 581 (1966).

\end{thebibliography}
\end{document}